\documentclass[conference]{IEEEtran}
\IEEEoverridecommandlockouts

\usepackage{amsmath,amssymb,amsfonts}

\usepackage{cite}
\usepackage{graphicx}
\usepackage{textcomp}
\usepackage[svgnames]{xcolor}
\usepackage{comment}
\usepackage{url}
\usepackage{listings}
\usepackage{booktabs}
\usepackage[a4paper, total={184mm,239mm}]{geometry}

\usepackage{makecell}
\usepackage{placeins}
\usepackage{hyperref}

\usepackage{algorithm}
\usepackage{algpseudocode}

\newcolumntype{L}[1]{>{\raggedright\arraybackslash}p{#1}}
\newcolumntype{C}[1]{>{\centering\arraybackslash}p{#1}}

\algrenewcommand\algorithmicrequire{\textbf{Require:}}
\algrenewcommand\algorithmicensure{\textbf{Ensure:}}
\algrenewcommand\algorithmiccomment[1]{\hfill$\triangleright$~#1}
\newcommand{\fun}[1]{\textsc{#1}}

\def\BibTeX{{\rm B\kern-.05em{\sc i\kern-.025em b}\kern-.08em
    T\kern-.1667em\lower.7ex\hbox{E}\kern-.125emX}}

\usepackage{caption}

\IEEEoverridecommandlockouts
\usepackage{enumitem}
\usepackage[utf8]{inputenc}
\usepackage{multirow}

\usepackage[absolute]{textpos}

\def\BibTeX{{\rm B\kern-.05em{\sc i\kern-.025em b}\kern-.08em
    T\kern-.1667em\lower.7ex\hbox{E}\kern-.125emX}}

\usepackage{tikz}
\usepackage{subfig}
\usepackage{fancyvrb}
\usepackage{array}

\usepackage[strings]{underscore}

\begin{document}

\newcommand{\papernote}{This paper will be presented at Design, Automation and Test in Europe Conference (DATE) 2026}

\title{%
  {\large \papernote\par}%
  \vspace{0.2em}%
  LAsset: An \underline{L}LM-assisted Security \underline{Asset} Identification Framework for System-on-Chip (SoC) Verification\\[-0.9em]
}

\author{
\IEEEauthorblockN{\small Md Ajoad Hasan, Dipayan Saha, Khan Thamid Hasan, Nashmin Alam,
Azim Uddin, Sujan Kumar Saha, Mark Tehranipoor, Farimah Farahmandi}
\IEEEauthorblockA{\textit{\small Department of Electrical and Computer Engineering, University of Florida, Gainesville, FL, USA} \\
\{\small md.hasan, dsaha, khanthamidhasan, nashminalam, azim.uddin, sujansaha\}@ufl.edu,
\{\small tehranipoor, farimah\}@ece.ufl.edu}
\vspace{-5mm}
}

\maketitle

\begin{abstract}
The growing complexity of modern system-on-chip (SoC) and IP designs is making security assurance difficult day by day. One of the fundamental steps in the pre-silicon security verification of a hardware design is the identification of security assets, as it substantially influences downstream security verification tasks, such as threat modeling, security property generation, and vulnerability detection. Traditionally, assets are determined manually by security experts, requiring significant time and expertise. To address this challenge, we present \textit{LAsset}, a novel automated framework that leverages large language models (LLMs) to identify security assets from both hardware design specifications and register-transfer level (RTL) descriptions. The framework performs structural and semantic analysis to identify intra-module primary and secondary assets and derives inter-module relationships to systematically characterize security dependencies at the design level. Experimental results show that the proposed framework achieves high classification accuracy, reaching up to ~90\% recall rate in SoC design, and ~93\% recall rate in IP designs. This automation in asset identification significantly reduces manual overhead and supports a scalable path forward for secure hardware development.
\end{abstract}


\section{Introduction}
\label{sec:introduction}

Modern system-on-chip (SoC) designs have grown into highly complex systems that integrate processors, memory, bus, and a wide range of peripherals onto a single chip \cite{rowen2008engineering, Chakravarthi2023}. While this high degree of integration enhances performance and cost-effectiveness, it simultaneously introduces new security challenges by significantly expanding the attack surface \cite{8000621, kocher2011complexity}. At the same time, increasing market pressure often forces shorter development cycles, leaving limited room for comprehensive pre-silicon security verification. As a result, many vulnerabilities remain undetected until the post-silicon phase \cite{9116483, pearce2023high}, where mitigation is substantially more resource-intensive \cite{mitra2010post}. To ensure robust protection, it is therefore critical to address security issues as early as possible in the design cycle \cite{arm2009arm}. Early identification of security-critical components not only reduces the risk of costly fixes later in the flow but also builds confidence in the overall design trustworthiness.

\begin{figure}[t]
\centering
\includegraphics[scale=.4]{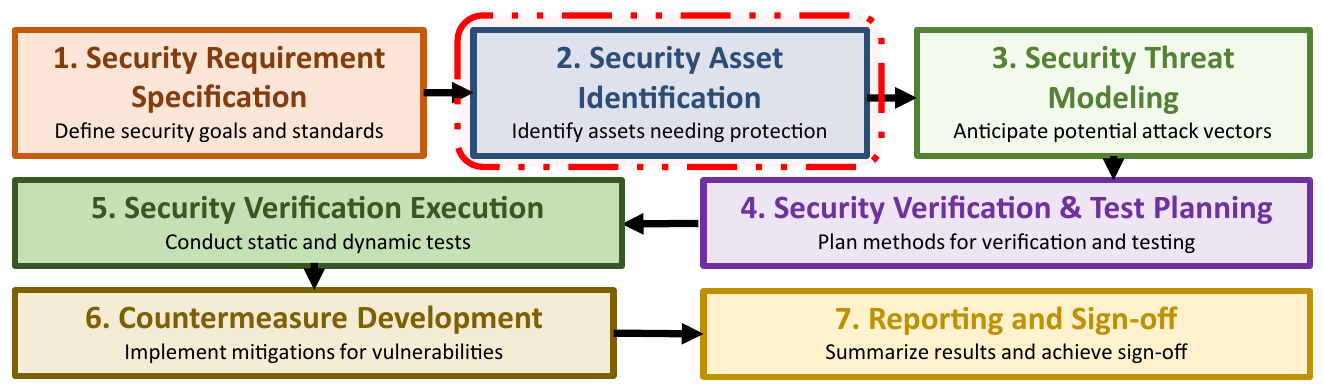}
\caption{Security Verification Flow in hardware design}
\label{flow}
\vspace{1mm}
\end{figure}

In reliable and secure hardware systems design, accurately identifying security assets early on provides the foundation for subsequent tasks, such as threat modeling, test plan generation, security verification, and countermeasure development, as shown in Figure \ref{flow} \cite{portillo2016building, slpsk2023protects, contreras2017security, peeters2015soc}. As underscored by \textit{Security Annotation for Electronic Design Integration (SA-EDI)} standard \cite{b4}, asset identification is not merely the initial but arguably the most important step toward sustainable security assurances in hardware systems. Despite its widely recognized significance, asset identification has traditionally remained a manual process based on industry expertise \cite{meza2023information, 7880823}, making it not only inefficient and error-prone but also difficult to scale for large SoC designs. Only a few recent works have explored asset identification, and those that do often consider it in a limited context. For instance, the SAIF tool \cite{farzana2021saif} attempts to identify assets in hardware designs by analyzing vulnerabilities and threat models, while the LASHED flow \cite{ahmad2025lashed} leverages common weakness enumeration (CWE) vulnerabilities for the same purpose. However, both approaches identify assets only after analyzing design vulnerabilities, whereas asset identification should be the primary step in a security verification flow, as illustrated in Figure 1. Similarly, the method in \cite{nath2025toward} relies heavily on RTL signal and register naming conventions, using pattern matching against known security-critical identifiers; as a result, it fails to generalize to designs that do not follow such conventions and offers no reasoning for why the detected elements should be considered assets.

Motivated by the challenges in today’s asset identification approaches, we outline the following research questions (\textbf{RQs}):

\underline{\textbf{RQ 1:}} How can we determine which elements are considered security-relevant assets within a given hardware design architecture?

\underline{\textbf{RQ 2:}} Can we develop an end-to-end automated methodology for reliably identifying security assets across diverse SoC and IP designs?

\underline{\textbf{RQ 3:}} How can we validate that an identified asset is really an asset?

\begin{figure}[t]
\centering
\includegraphics[scale=.433]{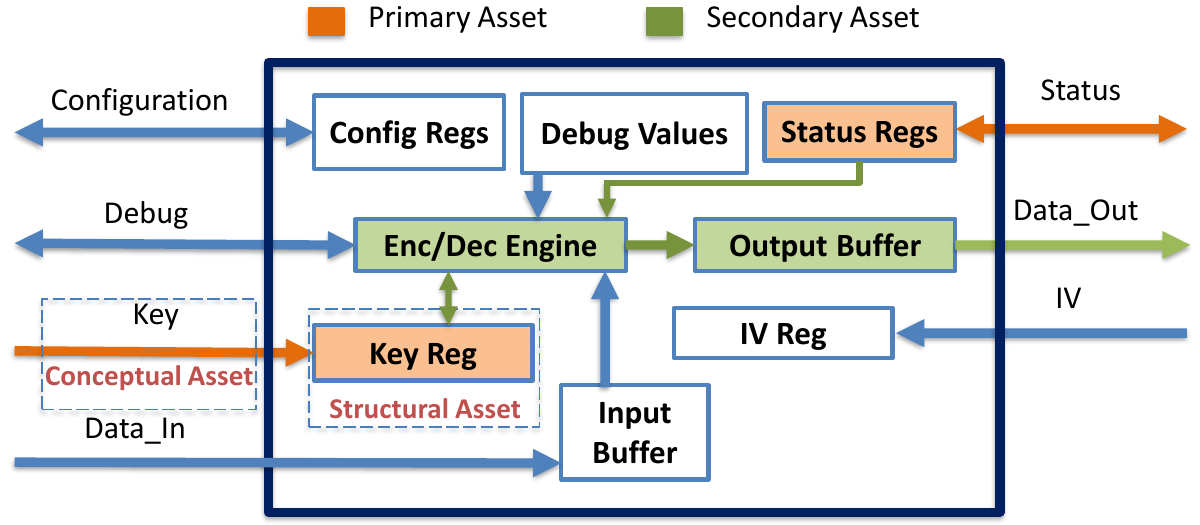}
\caption{Overview of Asset Types in an AES Encryption Design}
\label{fig1}
\vspace{1mm}
\end{figure}

After investigating the answers to the above questions, we propose \textit{LAsset}, the first-ever LLM-assisted automated framework for comprehensive security asset identification in SoC and IP designs, motivated by the success of LLM-assisted frameworks in hardware security verification \cite{sahaaccess,saha2025sv,socurellm,11022932,bugwhisperer,saha2024empowering}. \textit{LAsset} leverages the SA-EDI standard and IEEE P3164 guidelines for preliminary identification of assets using LLM. It further validates and refines the asset list by mapping it against the CWE vulnerability database, thereby ensuring its trustworthiness. In summary, the key contributions of our work are as follows:

\begin{itemize}
\item \textbf{Novelty:} In this paper, we present \textit{LAsset}, the first framework to leverage LLMs for the automated identification of security assets from SoC and IP designs.

\item \textbf{Robustness and Versatility:} \textit{LAsset} operates with either combined specification and RTL inputs or RTL alone, supports Verilog, SystemVerilog, and VHDL, and applies to both IP and SoC designs, ensuring broad applicability.

\item \textbf{Comprehensiveness:} Beyond asset identification, \textit{LAsset} associates each asset with threat modeling details and security justifications, providing actionable context for downstream verification tasks.

\item \textbf{Trustworthiness:} A refinement stage validates each asset by cross-referencing the CWE database and applying self-consistency checks, yielding a high-fidelity asset list with up to 93\% recall on real designs.
\end{itemize}

The rest of this paper is organized as follows. Section ~\ref{sec:background} provides a brief account of the background and rationale. Section ~\ref{sec:methodology} describes the proposed methodology. The experimental results and comparative analysis are presented in Section ~\ref{sec:experimental} before concluding the paper in Section ~\ref{sec:conclusion}.

\section{Background}
\label{sec:background}
\subsection{Security Asset}
By definition, a security asset in SoCs is any hardware component or data element whose protection is essential to preserve the system’s confidentiality, integrity, or availability (CIA). Security assets can be classified based on two criteria: (i) abstraction and (ii) dependency. According to IEEE P3164 \cite{b5}, Security assets can be classified as conceptual and structural depending on the abstraction level of the design.

\begin{description}[leftmargin=*, labelindent=0.5mm, labelsep=1.5mm]
    \item[$\bullet$] \textbf{Conceptual Assets}:
    Conceptual Assets are high-level information elements tied to the system’s use-case flows. For example, as shown in Figure \ref{fig1}, in an AES encryption engine, the encryption key is a conceptual asset because its confidentiality must be preserved regardless of where it resides in the design.

    \item[$\bullet$] \textbf{Structural Assets}: 
    Structural Assets are the hardware elements that physically store or carry conceptual assets. These include registers, buffers, latches, or gates involved in handling or storage of conceptual assets. In the same AES engine, the Key Register is a structural asset because it directly stores the encryption key value.
\end{description}

In addition, security assets can be categorized into Primary and Secondary Assets based on their dependency role in potential security breaches \cite{farzana2021saif}.

\begin{description}[leftmargin=*, labelindent=0.5mm, labelsep=1.5mm]
    \item[$\bullet$] \textbf{Primary Assets}:
    Primary Assets are components that serve as the direct target of an attack. Continuing the AES example, the encryption key itself is the primary asset since attackers seek to compromise it directly.
    
    \item[$\bullet$] \textbf{Secondary Assets}:
    Secondary Assets are components that interact with or facilitate the exposure of primary assets. While not directly critical, their compromise can weaken the overall security of the SoC by retrieving the associated primary asset. For example, system buses, peripheral ports, and internal signals/registers that carry the data of the primary asset, either fully or partially. In AES, elements such as the Encryption/Decryption Engine and the Output Buffer qualify as secondary assets because, though they are not the final target, their compromise can indirectly expose the encryption key.
\end{description}

\begin{figure}[t]
\centering
\includegraphics[scale=.243]{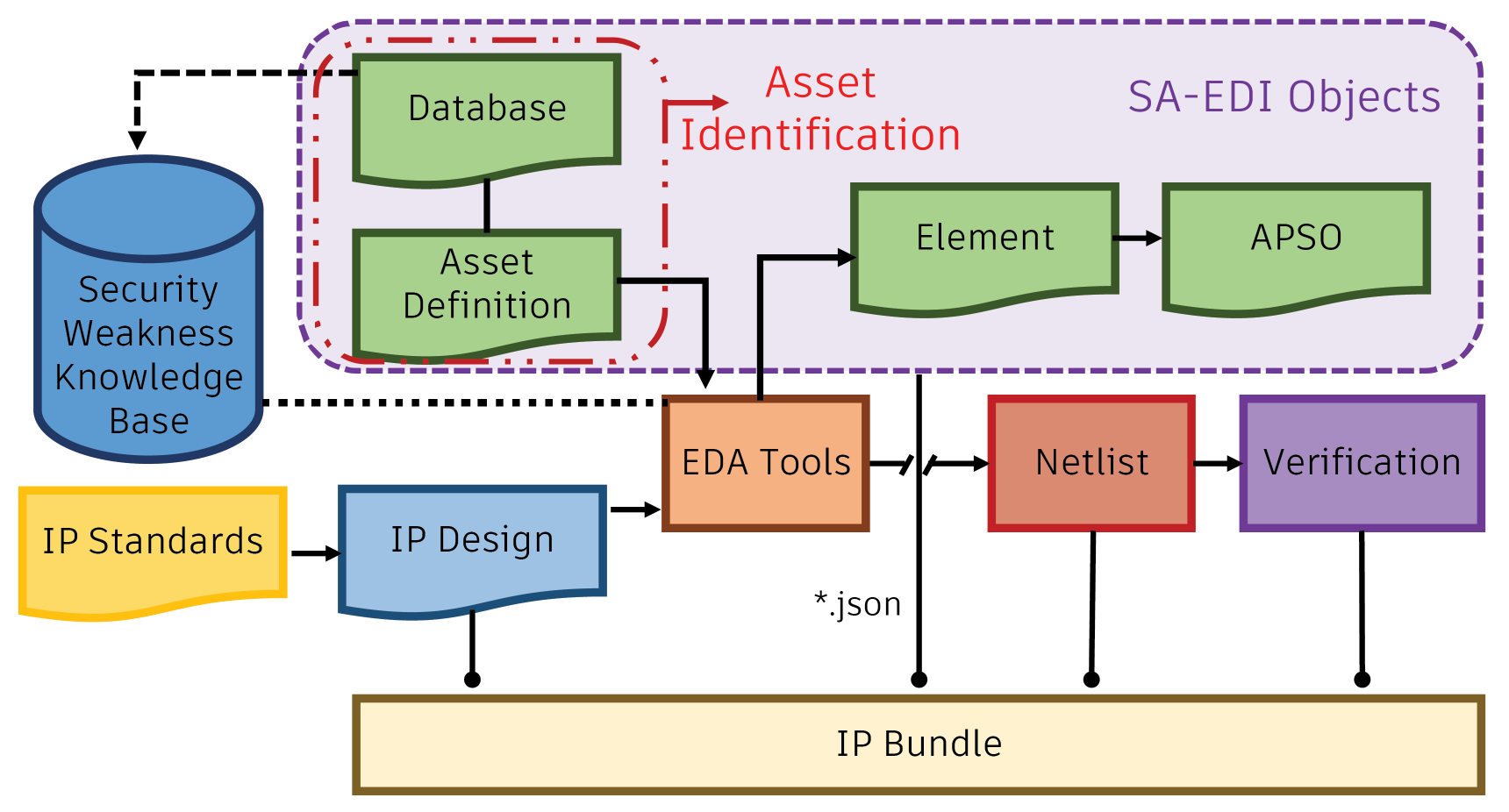}
\caption{\textit{SA-EDI} IP bundle \cite{b4}}
\label{saedi}
\vspace{1mm}
\end{figure}

\vspace{-3mm}
\subsection{Security Annotation for Electronic Design Integration (SA-EDI) standard}

\begin{figure*}[t]
\centering
\includegraphics[scale=.61]{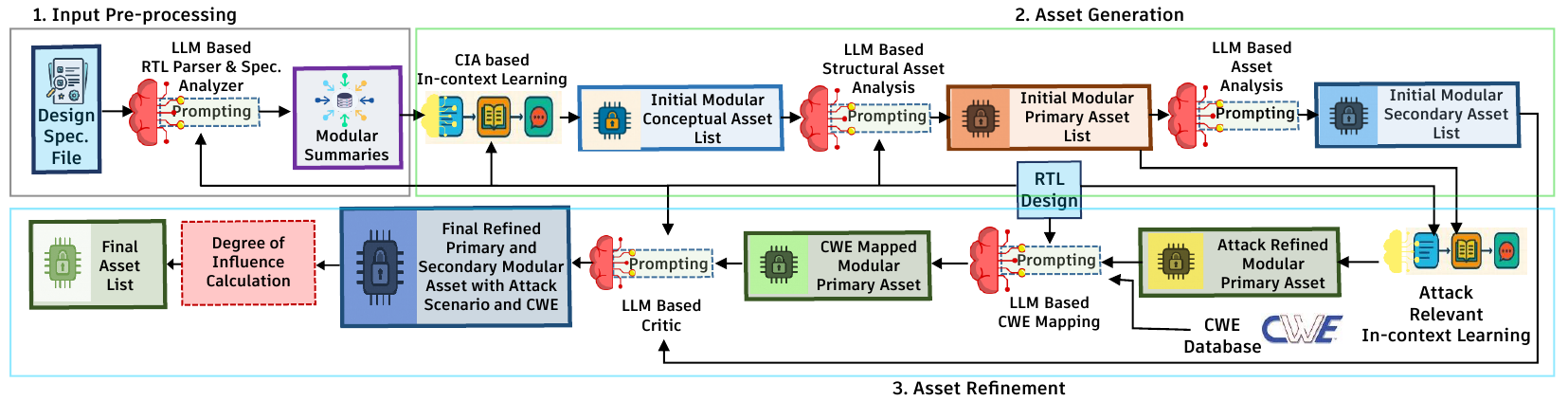}
\caption{Overview of the proposed \textit{LAsset} framework for security asset identification}
\label{frame}
\vspace{-6mm}
\end{figure*}

The SA-EDI standard provides specification guidelines for documenting the security concerns of hardware IPs as they are integrated into SoCs. As illustrated in Figure \ref{saedi}, security-related information—including asset definitions, security weakness databases, port- and parameter-related elements, and Attack Points Security Objectives (APSO)—is captured alongside the IP design and represented in a SA-EDI object in JSON format. The standard emphasizes that the first step is to identify conceptual assets in the design that require protection under security objectives such as confidentiality, integrity, and availability (CIA), along with their associated structural elements (e.g., ports, parameters) that could compromise these objectives. To support this process, the standard refers to the IEEE P3164 white paper \cite{b5}, which introduces the \textit{Conceptual and Structural Analysis (CSA)} methodology. \textit{CSA} begins by identifying conceptual assets tied to the CIA triad and then mapping them to their corresponding RTL representations, i.e., structural assets that are passed to the next Asset Definition objects in SA-EDI. This methodology highlights the distinction between conceptual assets—the 'what' that requires protection, and structural assets—the 'where' in the design that embodies that asset—ensuring that IP developers document security-critical assets comprehensively before integrating IPs into larger SoC designs. Accordingly, to address our \textbf{RQ1} in Section \ref{sec:introduction}, we build on the insights provided in the SA-EDI guidelines.

\vspace{1mm}

\section{LAsset Framework}
\label{sec:methodology}

Identifying security assets as posed in \textbf{RQ1-RQ3}, mentioned in Section \ref{sec:introduction}, requires a solution with several key qualities. Following the SA-EDI principle of first defining conceptual assets based on CIA objectives and then mapping them to structural design elements, it must understand both natural-language specification and RTL to map conceptual assets to their structural counterparts; reason over design semantics, hierarchy, and inter-module dependencies to generalize across diverse SoC and IP designs; and provide verifiable, standards-aligned justifications to ensure the fidelity of identified assets. Such a task requires semantic comprehension, contextual reasoning, modular task decomposition, and explainable decision-making capabilities that traditional rule-based approaches lack. LLMs inherently offer these properties: they can jointly interpret specification and RTL, infer security relevance from context rather than naming heuristics, be organized into specialized agents for sub-tasks like specification analysis, RTL analysis, asset mapping, and CWE-based rationalization, and generate human-readable rationales with self-consistency checks. Leveraging these strengths, we design \textit{LAsset} as an agentic framework driven by an LLM capable of performing automated, scalable, and reliable asset identification.

As shown in Figure \ref{frame}, \textit{LAsset} operates through three sequential agents: (i) \textit{Input Pre-processing}, which parses the available specification and RTL resources and aligns them under the SA-EDI-based conceptual–structural asset model; (ii) \textit{Asset Generation}, where specialized LLMs analyze the pre-processed inputs to identify candidate conceptual assets and map them to their structural counterparts; and (iii) \textit{Asset Refinement}, which validates and rationalizes each candidate through cross-referencing with the CWE database and self-consistency checks, producing a final, standards-aligned asset list. This modular agentic design enables \textit{LAsset} to perform automated, scalable, and verifiable security asset identification across diverse SoC and IP designs.

\subsection{Input Pre-processing}
Between the two inputs of \textit{LAsset} framework, the RTL repository provides the internal functionalities as well as information flow tracking, whereas the corresponding Spec.\ primarily outlines the module’s integration within the broader SoC architecture, providing additional design insights and interface details in natural language. Therefore, we pre-process the inputs in terms of the security concerns of the design so that LLMs can better understand the true security context rather than hallucinating due to token limitations and constrained long-context retention. The \textit{Input Pre-processing} agent performs the following three major tasks, as shown in Algorithm \ref{alg:lasset} (line 3-4):

\subsubsection{Design Modules Extraction}
In this task, LLMs' reasoning is employed to (i) prune non-relevant modules and (ii) select security-critical modules among the extracted module names from the RTL repository. For example, within the NEORV32 SoC, the \textit{neorv32\_application\_image} IP module is not considered security-critical and was therefore excluded during this task, whereas the security-sensitive \textit{neorv32\_cpu\_pmp} IP module, among others, was retained. This task is important, for it makes the flow much more resource-efficient (both cost and time) by reducing the security non-relevant design overhead in the beginning.

\subsubsection{Modular Spec.\ Summarization}
This task is built on a Retrieval-Augmented Generation (RAG) in-context learning-based model, where the SoC-level design Specs. is used as the knowledge source. For each security-relevant module, a user-guided query augments the prompt to produce a \emph{Technical Summary} via LLMs, which contains all the security-relevant information both within the module and across modules in the broader SoC context. Overall, this \emph{summary} acts as a concise Spec. input for the \textit{Asset Generation} agent for the respective module.

\subsubsection{Modular RTL Parsing}
To extract the design elements, we implement two LLM-based parsers: one for I/O ports and another for internal signals/registers, along with their types and functions at the module level. This is done to ensure that LLMs do not overlook any of the parsed design elements for asset decision, thereby reducing the chance of false negatives.

\subsection{Asset Generation}
Using the \textit{Technical Summary} and RTL as input, we employ the few-shot in-context learning paradigm for asset generation. We define the hardware security context, supplemented by the Q/A related to CIA security objectives. To further illustrate the task, we provide several case studies featuring widely recognized hardware IP blocks, e.g., AES, GPIO, Gaussian Noise Generator, each annotated with asset listings and corresponding justifications explaining why some design elements are security assets and others are not. Through this step-by-step approach, the LLM is systematically guided to learn and identify security assets.

After this in-context learning, through a stepwise reasoning and appropriate constraints approach, we identify the conceptual assets at the module level. These assets are then systematically mapped to their corresponding structural RTL references, derived from the parsed design elements- these are the \textit{primary assets} at the module level. In addition, for each \emph{primary asset}, we find the internal signals/registers that influence/violate its security objective(s)- these are termed as the \emph{secondary assets} at the module level. \textit{Asset Generation} steps are shown in Algorithm \ref{alg:lasset} (line 5-6).

\vspace{-2mm}

\subsection{Asset Refinement}
When applied to security asset identification, LLMs often lean toward listing a broad set of possible security and non-security assets. This tendency, while helpful in maximizing coverage, can also introduce false positives by including design elements that are hardly security-relevant. For this reason as well as to answer our \textbf{RQ3} in section \ref{sec:introduction}, a rigorous three-stage back-to-back refinement agent is designed to filter out the false positives and validate the remaining results, ensuring that the final asset list at the module level is accurately substantiated. The refinement agent comprises the following stages, also shown in Algorithm \ref{alg:lasset} (line 7-17):

\subsubsection{Attack Scenario Analysis}
We evaluate each candidate primary asset for susceptibility to seven hardware attack classes: side-channel, fault injection, secure-to-nonsecure information leakage, unauthorized access, privilege escalation, hardware Trojan, and denial-of-service. Using in-context training, LLMs generate high-level abstractions of plausible attack scenarios for each asset; assets without such scenarios are excluded. This analysis justifies asset inclusion to verification experts and, through the zero-shot reasoning capabilities of LLMs, extends beyond modular boundaries to capture system-wide security implications.

\subsubsection{CWE Mapping}
To provide additional context for the generated assets as well as refine the assets further, we analyze whether and to what extent the CWEs from the MITRE database can be mapped to each primary asset \cite{b6}. Given the substantial size of the database, the analysis is structured to examine CWEs under each security objective one at a time, ensuring that no CWE is overlooked. Besides, we focus on only the security asset-relevant information in the database to prevent LLMs from being overwhelmed by extraneous information during the CWE mapping process. The rigorous mapping ensures that only the absolutely critical CWE(s), if any, are mapped to each asset. Finally, Assets with no mappable CWE are removed. 

\subsubsection{Self-critique Revision}
The asset annotations generated at the module level so far are challenged by a self-critique prompt and revised further. Using the input resources again, we re-check asset relevance, RTL references, security-objective correctness, attack scenario coherence, and CWE appropriateness, thereby improving both accuracy and reliability.

\begin{algorithm}[t]
 
\scriptsize                                
\caption{\textit{LAsset} Framework}
\label{alg:lasset}

\begin{algorithmic}[1]

\Require Design Spec $S$; RTL repo $R=\{m_1,m_2,m_3,\ldots\}$; \texttt{DBASEcwe};
\Require Function Library: \texttt{SpecRAG(Spec)}, \texttt{SecAsset(Asset, Signals, RTL)}, \texttt{UnrollAsset(Asset_List, Path)}, \texttt{ICLasset}, \texttt{ICLattack};
\Require Prompt Library: \texttt{LLMparse}, \texttt{LLMasset}, \texttt{LLMattack}, \texttt{LLMcwe}, \texttt{LLMref}.
\Ensure Group of modular asset annotations $\texttt{Asset}_g=\{\texttt{Ref\_Asset}_m \mid m\in M\}$.
\Ensure List of DoI per secondary asset $\texttt{DOI}=\{\texttt{DoIsec\_asset}_h \mid h\in H\}$

\State \(M \gets \textsc{Mod\_listing}(R)\)

\Statex \textbf{Step: Modular Assets Identification}
\For{each \(m \in M\)}
  \Statex \texttt{/* Input pre-processing */}
  \State \textsc{Mod\_Spec} \(S_m \gets \fun{SpecRAG}(S)\)
  \State \textsc{Mod\_RTLparse} \(\{\mathrm{Prm}_m,\mathrm{Sec}_m\} \gets \fun{LLMparse}(R(m))\) 
  \Statex \texttt{/* Assets generation */}
  \State \(\mathrm{Asset}_m \gets \fun{LLMasset}(S_m,\mathrm{Prm}_m,R(m),\fun{ICLasset})\)
  \State \(\mathrm{Exp\_Asset}_m \gets \fun{SecAsset}(\mathrm{Asset}_m,\mathrm{Sec}_m,R(m))\)
  \Statex \texttt{/* Assets refinement */}
  \State \(\mathrm{Ref\_Asset\mbox{-}1}_m \gets \fun{LLMattack}(\mathrm{Exp\_Asset}_m,\fun{ICLattack})\)
  \State \(\mathrm{Ref\_Asset\mbox{-}2}_m \gets \fun{LLMcwe}(\mathrm{Ref\_Asset\mbox{-}1}_m,\mathrm{DBASEcwe})\)
  \State \(\mathrm{Ref\_Asset}_m \gets \fun{LLMref}(\mathrm{Ref\_Asset\mbox{-}2}_m)\)
\EndFor

\Statex \textbf{Step: DoI calculation}
\State $H \gets \textsc{hierarch\_path}(R)$

\For{path $h \in H$}
    \Statex \texttt{/* Enumerate linked modules' assets */}
    \State $\mathcal{A}_{h,g} \gets \textsc{UnrollAssets}(A_g, h)$
    \State \textbf{let } $\mathcal{A}_{h,g} \triangleq \{Amh_1, Amh_2, Amh_3, \ldots\}$  
    \State $\texttt{DoIsec\_asset}_h \gets \textsc{DoI}(\mathcal{A}_{h,g})$
    \State \textbf{let } $\texttt{DoIsec\_asset}_h \triangleq \{{DoIsec\_asset}_{h,1}, {DoIsec\_asset}_{h,2}, \ldots\}$

\EndFor

\State \Return \(\mathrm{Asset}_g\), \texttt{DOI}
\end{algorithmic}
\end{algorithm}
\setlength{\textfloatsep}{8pt}

\subsubsection{Degree of Influence (DoI) Calculation for Secondary Assets}
After enumerating assets at the module level, we assess how these assets influence one another within the design. As modular assets interact across boundaries, their interdependencies determine how local vulnerabilities may escalate into broader system risks. RTL analysis is used to extract hierarchical linkages among security-critical modules. Along each path, the most tamper-prone asset is designated as the \emph{primary} asset, while the remaining assets are treated as \emph{secondary}. Bit-level RTL connectivity is then analyzed to quantify the relationship between primary and secondary assets, and the influence is backtracked multiplicatively along each path to derive the final \textit{DoI} metric for each secondary asset.

\vspace{-2mm}

\begin{equation}
\textit{DoI} =
\frac{\begin{array}{c}
\text{\# of bits of the secondary asset} \\
\text{connected to the primary asset}
\end{array}}
{\text{Total \# of bits of the primary asset}}
\times 100\%
\end{equation}

We compute DoI as an empirical metric to quantify the hardness of the identified secondary assets, which can also guide subsequent research within this scope.

\vspace{1mm}

\section{Experimental Results and Evaluations}
\label{sec:experimental}
We present case studies on the NEORV32 RISC-V SoC and several hardware IP blocks, under two configurations, to demonstrate the efficacy and generality of the \textit{LAsset} workflow in identifying security assets across diverse architectures. Results are available online.\footnote{\url{https://github.com/Ajoad/LAsset-Security-Assets}}

\begin{figure}[htbp]
\centering
\includegraphics[scale=.395]{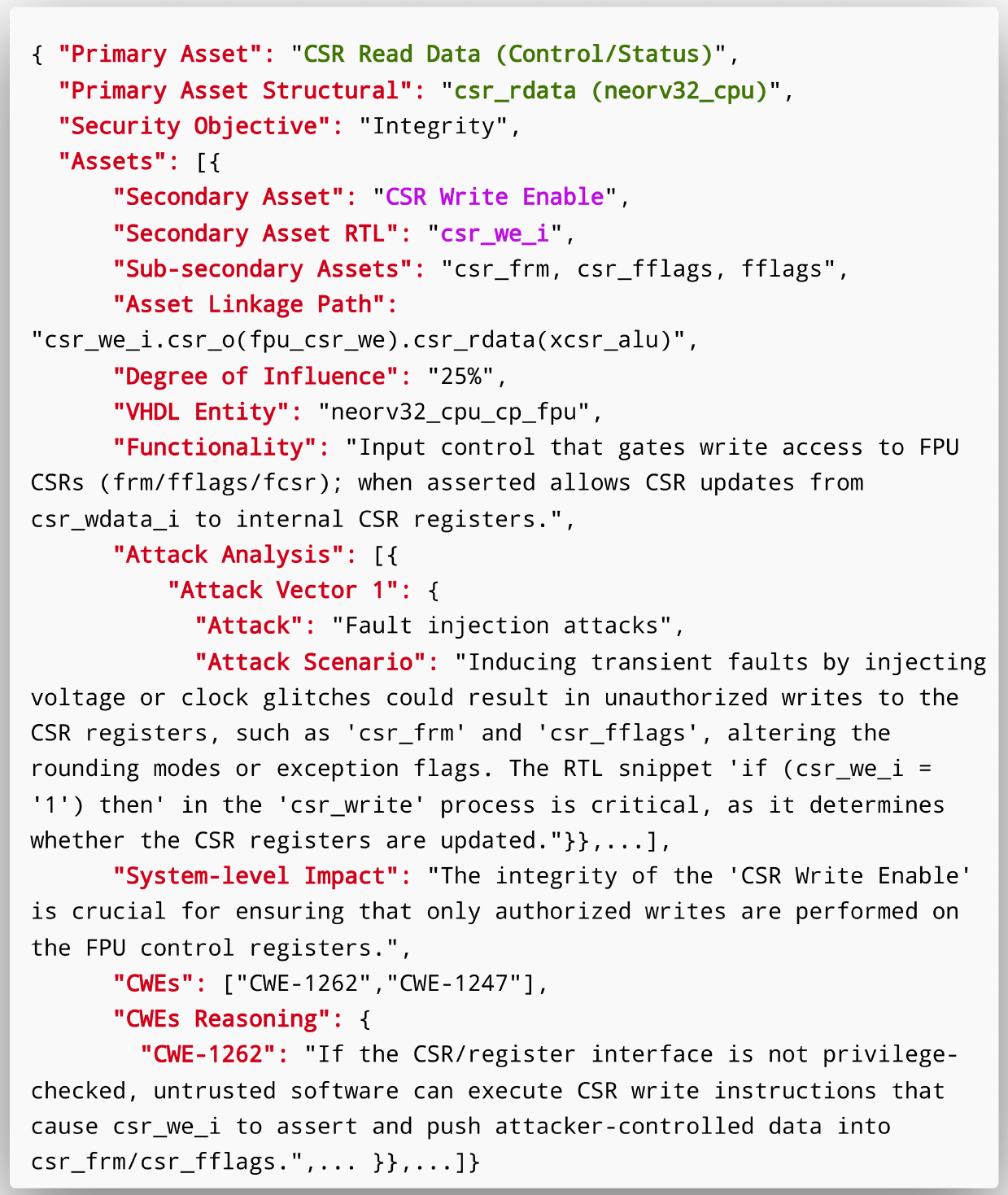}
\caption{A snippet of \textit{LAsset} output for NEORV32 processor (\textit{Spec. + RTL} approach)}
\label{fig7}
\vspace{-2mm}
\end{figure}

\textit{LAsset} employs OpenAI’s GPT-5 model for in-context learning–based asset identification. In the \textit{Modular Spec. Generation} task, retrieval is performed with 1,000-character chunks and a 200-character overlap to preserve context. Embeddings are generated using OpenAI’s \textit{text-embedding-ada-002} model, and similarity search is conducted with FAISS to retrieve the top 20 relevant chunks per query.

\vspace{-1mm}

\subsection{SoC: NEORV32 RISC-V processor}
For the SoC case study, we select the NEORV32 RISC-V processor for evaluating the \textit{LAsset} framework.

After a full \textit{LAsset} run, assets are identified across NEORV IPs, and their interconnections yield the \textit{Degree of Influence} of secondary assets on their primary counterparts. For example, the primary asset \textit{csr_rdata} in \textit{neorv32_cpu} is linked to several secondary assets, including \textit{csr_we_i} in \textit{neorv32_cpu_cp_fpu}. This asset controls only 8 of 32 FPU CSR bits (\textit{csr_fflags[4:0]}, \textit{csr_frm[2:0]}), giving $8/32=25\%$. As \textit{csr_o} maps fully to \textit{csr_rdata}, the overall influence remains 25\%. Associated threats include fault injection, unauthorized access, and hardware Trojans (CWE-1262, CWE-1247).

To validate the outputs, we manually create a reference list of security assets for each NEORV32 IP, based on our knowledge of hardware security assurance. We further cross-validate the manually annotated assets using mainstream LLM chatbots, i.e., GPT, Gemini, and Grok. This is carried out to assess the reliability of the manual annotations against the ground truth, while also introducing an additional evaluation dimension beyond human review, which might be subject to error. We compare these manual annotations with the responses from \textit{LAsset} whose summary is shown in Table III. Since \textit{LAsset} deterministically computes \textit{DoI} based on the relationships among the already validated primary and secondary assets, we do not perform a separate validation for this.

\begin{table}[htbp]
\centering
\tiny
\caption{Summary of Assets for Spec.+RTL vs RTL approaches across IP repositories. Each block reports Reference Asset List (Golden), Total Design Elements (Elem.), Total Initial Assets after \textit{Asset Generation} (Init.), Total Refined Assets after \textit{Asset Refinement} (Ref.), True Positive Assets after comparing with Nath et. al (TP), False Negative Assets (FN), and False Positive Assets (FP).}
\label{tab:asset_summary}
\setlength{\tabcolsep}{3pt}
\renewcommand{\arraystretch}{1.05}
\begin{tabular}{c c c c ccccc ccccc}
\toprule
\multirow{3}{*}{\textbf{Repository}} & \multirow{3}{*}{\textbf{IP}} & \multirow{3}{*}{\textbf{Golden}} & \multirow{3}{*}{\textbf{Elem.}} 
& \multicolumn{5}{c}{\textbf{Spec + RTL}} & \multicolumn{5}{c}{\textbf{RTL}} \\
\cmidrule(lr){5-9} \cmidrule(lr){10-14}
 & & & & Init. & Ref. & TP & FN & FP & Init. & Ref. & TP & FN & FP \\
\midrule
\multirow{3}{*}{OpenTitan}
  & hmac        & 11 & 268  & 15 & 11 & 8  & 3 & 3 & 14 & 12 & 8  & 3 & 4 \\
           & keymgr      & 42 & 688  & 46 & 44 & 37 & 5 & 7 & 44 & 42 & 36 & 6 & 6 \\
           & ot\_gpio    & 5  & 114  & 8  & 7  & 5  & 0 & 2 & 7  & 7  & 5  & 0 & 2 \\
\midrule
\multirow{6}{*}{OpenCores}
  & sha3        & 7  & 71   & 13 & 12 & 7  & 0 & 5 & 10 & 8  & 7  & 0 & 1 \\
           & simple\_gpio & 2 & 17   & 3  & 3  & 2  & 0 & 1 & 5  & 4  & 2  & 0 & 2 \\
           & tiny\_aes   & 4  & 141  & 10 & 8  & 4  & 0 & 3 & 13 & 13 & 4  & 0 & 8 \\
           & ahb\_m\_wishbone & 7 & 31 & 12 & 9 & 7 & 0 & 2 & 9  & 8  & 6  & 1 & 2 \\
           & rc4         & 3  & 12   & 4  & 4  & 3  & 0 & 1 & 5  & 4  & 3  & 0 & 1 \\
           & robust\_axi2apb & 10 & 87 & 15 & 13 & 10 & 0 & 3 & 11 & 11 & 9  & 1 & 2 \\
\midrule
\multirow{5}{*}{NEORV32}
    & CPU         & 14 & 35   & 15 & 14 & 12 & 2 & 2 & 15 & 13 & 11 & 3 & 2 \\
           & PMP (CPU)   & 6  & 23   & 8  & 6  & 5  & 1 & 1 & 6  & 6  & 4  & 2 & 2 \\
           & Debug Transport & 5 & 25  & 7  & 7  & 5  & 0 & 2 & 8  & 6  & 4  & 1 & 2 \\
           & TRNG        & 7  & 39   & 9  & 8  & 6  & 1 & 2 & 7  & 7  & 5  & 2 & 2 \\
           & UART        & 10 & 44   & 12 & 11 & 10 & 0 & 1 & 9  & 9  & 8  & 2 & 1 \\
\midrule
\textbf{Total} &          & 133 & 1595 & 177 & 157 & 121 & 12 & 35 & 163 & 150 & 112 & 21 & 37 \\
\bottomrule
\end{tabular}
\end{table}

\vspace{-1mm}

\subsection{IP: OpenTitan and OpenCores IPs}
For validating \textit{LAsset} at IP-level security asset identification, we have selected a total of 21 open-source crypto IPs, Interface-GPIO IPs, and Interface-peripheral IPs, from OpenTitan \cite{b9}, OpenCores \cite{b10}, and other repositories. A comprehensive primary-secondary classified assets listing for a 128-bit AES architecture is shown in Table~\ref{tab:aes_assets}. We compare our asset outputs with those from Nath et. al \cite{nath2025toward}, as shown in Table~\ref{tab:conf_soc_ip}, since both their tool-generated list of primary assets and the benchmark datasets for the same IPs are publicly available \cite{b11}.

\begin{table}[htbp]
\centering
\caption{Security Assets for AES-128 Design}
\label{tab:aes_assets}
\setlength{\tabcolsep}{3pt}
\resizebox{\columnwidth}{!}{%
\begin{tabular}{C{2.1cm} C{2.1cm} C{2.4cm} C{2.9cm} C{4.4cm}}
\toprule
\multicolumn{2}{c}{\textbf{Primary Asset}} &
\multirow{2}{*}{\textbf{Security Objective}} &
\multirow{2}{*}{\makecell[c]{\textbf{Secondary Asset}\\\textbf{(Design Module)}}} &
\multirow{2}{*}{\textbf{CWE-ID}} \\
\cmidrule(lr){1-2}
\makecell[c]{\textbf{Conceptual}\\\textbf{Asset}} &
\makecell[c]{\textbf{Structural Asset}} & & & \\
\midrule

\multirow{5}{*}{\parbox[t]{2.1cm}{\centering Key Expansion\\ Logic Output}}
& \multirow{5}{*}{key}
& \multirow{5}{*}{Confidentiality}
& out-1 (expand-key-128) & 1300, 1191, 1239, 1258 \\
\cmidrule(lr){4-5}
& & & out-2 (expand-key-128) & 1300, 1191, 1313 \\
\cmidrule(lr){4-5}
& & & key (one-round)        & 1300, 1247, 1191, 1239, 1258, 1263 \\
\cmidrule(lr){4-5}
& & & key (final-round)      & 1300, 1247, 1191, 1239, 1258, 1263 \\
\cmidrule(lr){4-5}
& & & out (AES-128)          & 1247, 1319, 1384 \\
\midrule

\multirow{3}{*}{\parbox[t]{2.1cm}{\centering Final Encrypted\\ Output}}
& \multirow{3}{*}{out}
& \multirow{3}{*}{Availability}
& state-out (one-round)  & 1247, 1261, 1313 \\
\cmidrule(lr){4-5}
& & & state-out (final-round) & 1247, 1261, 1313 \\
\cmidrule(lr){4-5}
& & & state & 1300, 1191, 1258 \\
\midrule

\multirow{5}{*}{\parbox[t]{2.1cm}{\centering Initial Data\\ Processing State}}
& \multirow{5}{*}{state}
& \multirow{5}{*}{Confidentiality}
& state-in (one-round)   & 1300, 1247, 1323, 1313, 1263 \\
\cmidrule(lr){4-5}
& & & state-in (final-round) & 1300, 1247, 1323, 1313, 1263 \\
\cmidrule(lr){4-5}
& & & state (AES-128)        & 1300, 1191, 1258 \\
\cmidrule(lr){4-5}
& & & state (table-lookup)   & 1300, 1247, 1319 \\
\cmidrule(lr){4-5}
& & & P0, P1, P2, P3 (table-lookup) & 1300, 1247, 1319 \\
\bottomrule
\end{tabular}}
\end{table}

\vspace{-1mm}

\subsection{Performance Comparison and Analysis}
\begin{figure}[htbp]
\centerline{\includegraphics[scale=.535]{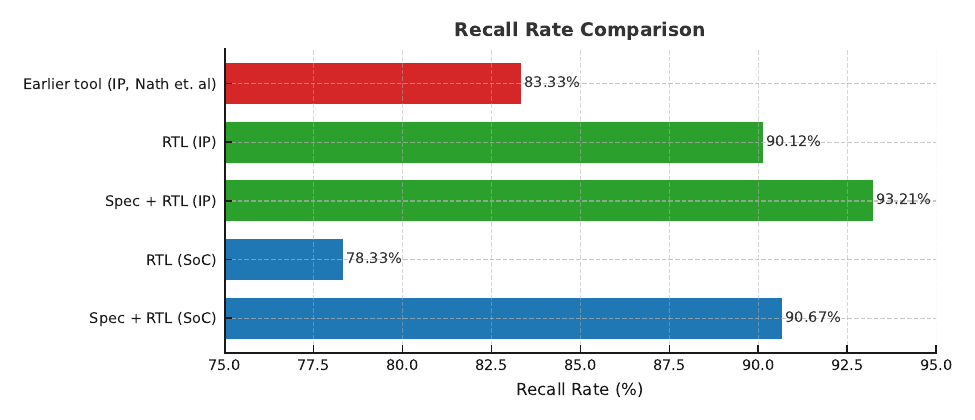}}
\caption{Graphical Comparison between \textit{RTL + Spec.} and \textit{Only RTL} Approaches at SoC and IP level}
\label{fig2}
\vspace{-2mm}
\end{figure}

In this sub-section, we compare the performance of \textit{LAsset}, as shown in Figure \ref{fig2}, to evaluate whether it aligns with the expected behavior and to analyze the underlying reasons for any observed discrepancies. A detailed, IP-wise summary of the results obtained by comparing the asset outputs along with the reference list, are presented in Table~\ref{tab:asset_summary}.

\subsubsection{\textbf{SoC}}

From Table~\ref{tab:conf_soc_ip}, we find that the \textit{Spec. + RTL} approach yields better results than the \textit{Only RTL} approach in SoC-level. This signifies that RTL is the primary contributing resource, since RTL explicitly maps out all necessary design components and their functionalities, which LLMs can interpret effectively. However, the Spec.'s natural-language description of operational behavior and interconnectivity helps the LLM better understand the design, more so than using RTL alone. Therefore, the takeaway from this analysis is that although an industry-standard Spec. alone is not enough to identify the security assets in design, it can work as a supplementary factor to the \textit{LAsset} framework.

\subsubsection{\textbf{IP}}

Table~\ref{tab:conf_soc_ip} shows that the \textit{Spec.+RTL} approach outperforms the \textit{Only RTL} approach at the IP-level. Compared with the work of Nath et al. \cite{b11}, \textit{LAsset} achieves a higher recall value of 93.21\%, whereas their approach has 83.33\%. These values are measured against their manually tailored golden asset list. Upon closely reviewing this list, however, we found that a number of actual assets are missing. For instance, in the case of the SHA-3 IP block, the round constants \texttt{rc1} and \texttt{rc2} should be included, as tampering with them would err the transformation steps and undermine the integrity of the algorithm. Because the golden list in Nath et al. \cite{b11} overlooks a significant number of such critical assets, those identified by \textit{LAsset} are incorrectly flagged as false positives, even though they must be true positives. Hence, a direct comparison of accuracy and F-1 score between \textit{LAsset} and Nath et al. \cite{b11} would not be a meaningful assessment; instead, recall rate provides a more appropriate metric for comparison.

\begin{table}[htbp]
\centering
\caption{Comparison across Different Input Configurations at SoC- and IP-levels.}
\label{tab:conf_soc_ip}
\tiny
\begin{tabular}{c c c c c}
\toprule
\textbf{Input Config} & \textbf{Actual Class} & \textbf{Predicted Positive} & \textbf{Predicted Negative} & \textbf{Performance Metric} \\
\midrule
\multicolumn{5}{c}{\textbf{SoC-level (NEORV32)}} \\
\midrule
\multirow{2}{*}{RTL + Spec.} & Positive & 272 & 30  & \multirow{2}{*}{Accuracy = 93.75\%} \\
\cmidrule(lr){2-4}
 & Negative & 59 & 1029 & \\
\midrule
\multirow{2}{*}{RTL} & Positive & 235 & 67  & \multirow{2}{*}{Accuracy = 91.16\%} \\
\cmidrule(lr){2-4}
 & Negative & 59  & 992 & \\
\midrule
\multicolumn{5}{c}{\textbf{IP-level (21 IPs from OpenTitan, OpenCores, etc.)}} \\
\midrule
\multirow{2}{*}{RTL + Spec.} & Positive & 148   & 15  & \multirow{2}{*}{Recall = 93.21\%} \\
\cmidrule(lr){2-4}
 & Negative & 50  & 1735  & \\
\midrule
\multirow{2}{*}{RTL} & Positive & 143  & 20  & \multirow{2}{*}{Recall = 90.12\%} \\
\cmidrule(lr){2-4}
 & Negative & 50  & 1735  & \\
\bottomrule
\end{tabular}
\vspace{-2mm}
\end{table}

\subsection{Time-Cost-Performance Comparison}

\begin{figure}[htbp]
\centerline{\includegraphics[scale=.29]{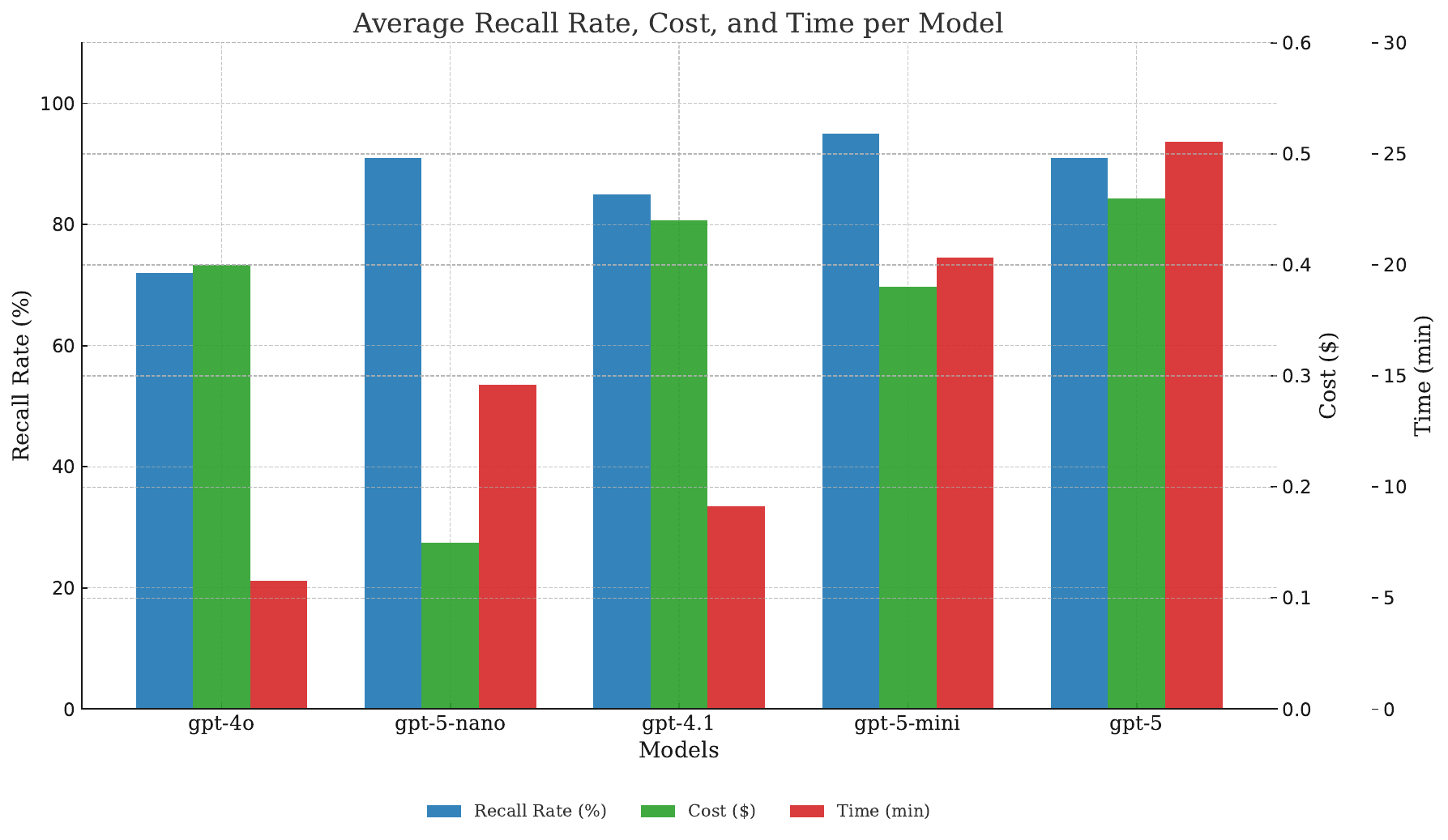}}
\caption{Performance-Time-Cost Comparison among different LLM models for \textit{LAsset}}
\label{fig9}
\vspace{-2mm}
\end{figure}

Although GPT-5 model has been employed to obtain the results of \textit{LAsset} presented in this paper, we consider the framework’s model-agnostic behavior as well. To this end, we executed the end-to-end flow for five sample IP blocks, i.e, AES, SHA-3, GPIO, AXI-Adapter, and AHB3LITE, and compared three outcomes: recall(\%), runtime, and monetary cost, among five different LLM models for each IP, as shown in Figure \ref{fig9}.

To quantize the trade-offs, we define a model-agnostic utility (MAU) that aggregates the normalized metrics under user-specified priorities:

\begin{equation}
\begin{aligned}
\mathrm{MAU} &= \alpha \frac{100 - R}{100}
               + \beta \frac{T}{T_{\max}}
               + \gamma \frac{C}{C_{\max}}, \\[6pt]
\text{where}\;\; & \alpha,\beta,\gamma \geq 0,\;\; \alpha + \beta + \gamma = 1.
\end{aligned}
\label{eq:mau}
\end{equation}

Here, $T_{\max}$ and $C_{\max}$ denote the maximum observed time and cost for a design; $R$, $T$ and $C$ respresents the observed recall percentage, time and cost for the corresponding LLM model, respectively; $\alpha$, $\beta$, and $\gamma$ indicate the selection preference (e.g., performance-centric, latency-sensitive, or cost-constrained). A lower $\mathrm{MAU}$ therefore corresponds to a more efficient model. For our experiments, we prioritize performance over cost and time, so we tune the equation \eqref{eq:mau} with $\alpha$ = 0.6, $\beta$ = 0.1, and $\gamma$ = 0.3. Under this configuration, GPT-5-nano is found to be the best candidate with the least (MAU).

\section{Conclusion}
\label{sec:conclusion}
This paper introduces \textit{LAsset}, an automated AI-assisted methodology for identifying security-critical assets in both hardware IPs and SoCs, drawing on the inherent capabilities of LLMs. \textit{LAsset} not only identifies security-relevant primary and corresponding secondary assets in hardware designs but also associates them with relevant attack vectors and CWEs, as well as the degree of influence per secondary asset, thus aiding design and verification engineers to better understand and utilize these assets in subsequent security stages. This development further standardizes existing IP and SoC design flows while enhancing the overall hardware security verification landscape.

\section*{Acknowledgment}
This research was funded by the U.S. National Science Foundation (NSF) under the Faculty Early Career Development (CAREER) Program, Grant No.~2339971.

\clearpage


\bibliographystyle{IEEEtran}
\bibliography{references}

\end{document}